\documentclass[journal]{IEEEtran}

\usepackage{amsmath}
\usepackage{amsfonts}
\usepackage{amssymb}
\usepackage{amsthm}
\usepackage{tabularx}
\usepackage{mathrsfs} 
\usepackage{breqn}
\usepackage{siunitx}
\usepackage[numbers,sort&compress]{natbib}

\usepackage{url}
\usepackage[hidelinks]{hyperref}

\newtheorem{remark}{Remark}

\newtheorem{lemma}{Lemma}

\usepackage{stfloats}
\usepackage{float}
\usepackage[font=small,skip=0pt]{caption}
\usepackage{graphicx}

\usepackage{subfigure}

\newcommand*{\hham}{^{\rm{H}}}

\usepackage[normalem]{ulem}

\makeatletter
\def\blfootnote{\xdef\@thefnmark{}\@footnotetext}
\makeatother

\usepackage[export]{adjustbox}

\usepackage{tikz}

\usepackage{pgfplots}
\pgfplotsset{compat=1.3}

\begin{document}
		\title{\huge{Improved SINR Approximation for Downlink \textcolor{black}{RSMA}-based Networks with Outdated Channel State Information}} 
        \author{Maria Cecilia Fern\'andez Montefiore,~Gustavo Gonz\'alez,~F. Javier~L\'opez-Mart\'inez,~and Fernando Gregorio
	}
 
	\maketitle
	\begin{abstract}
Understanding the performance of multi-user multiple-input multiple-output (MU-MIMO) systems under imperfect channel state information at the transmitter (CSIT) remains a critical challenge in next-generation wireless networks. In this context, accurate statistical modeling of the signal-to-interference-plus-noise ratio (SINR) is essential for enabling tractable performance analysis of multi-user systems.
This paper presents an improved statistical approximation of the SINR for downlink (DL) MU-MIMO systems with imperfect CSIT. The proposed model retains the analytical simplicity of existing approaches (e.g., Gamma-based approximations) while overcoming their limitations, particularly the underestimation of SINR variance.
We evaluate the proposed approximation in the context of Rate-Splitting Multiple Access (RSMA)-enabled MIMO DL systems with outdated CSIT. The results demonstrate excellent accuracy across a wide range of system configurations, including varying numbers of users, antennas, and degrees of CSIT staleness.

	\end{abstract}
	\maketitle
	\blfootnote{\noindent This work has been submitted to the IEEE for publication. Copyright may be transferred without notice, after which this version may no longer be accessible.} 
	\blfootnote{\noindent Manuscript received November 20, 2025; revised July 14, 2026. This work was supported in part by grant PID2023-149975OB-I00 (COSTUME) funded by MICIU/AEI/10.13039/501100011033 and FEDER/UE, in part by grant DGP\_PRED\_2024\_0212 funded by Consejer\'ia de Universidad, Investigaci\'on e Innovaci\'on of Junta de Andaluc\'ia and the European Union, and in part by grant PMA1-2025-14-017 funded by Asociaci\'on Universitaria Iberoamericana de Postgrado. The review of this paper was coordinated by XXXX. (\textit{Corresponding author: M.C. Fern\'andez Montefiore})}
	
	 	\blfootnote{\noindent M.C. Fern\'andez Montefiore and F.J. L\'opez-Mart\'inez are with the Dept. Signal Theory, Networking and Communications, Research Centre for Information and Communication Technologies (CITIC-UGR), University of Granada, 18071, Granada, Spain. E-mail:\texttt{\{mcfm,fjlm\}@ugr.es}. M.C. Fern\'andez Montefiore, G. Gonz\'alez and F. Gregorio are with Instituto de Investigaciones en Ingenier\'ia El\'ectrica ``Alfredo Desages'' (IIIE), UNS-CONICET, and Departamento de Ingenier\'ia El\'ectrica y de Computadoras, Universidad Nacional del Sur (UNS), Argentina. E-mail: \texttt{\{ggonzalez,fernando.gregorio\}@.uns.edu.ar}}

	\blfootnote{Digital Object Identifier 10.1109/XXX.2025.XXXXXXX}

\section{Introduction}\label{sec-intro}

The evolution of communication systems towards sixth-generation (6G) networks is motivated by the need to satisfy the growing demands for higher throughput, lower latency, enhanced reliability, heterogeneous quality of service (QoS), and massive connectivity \cite{Mao2022}. While fifth-generation (5G) networks based upon three key services -- enhanced mobile broadband (eMBB), ultra-reliable low-latency communications (URLLC), and massive machine-type communications (mMTC)-- 6G seeks to enhance and integrate these capabilities with emerging functionalities such as communication and sensing, localization, and computing \cite{Mishra2022,Clerckx2024}. Mobility scenarios present significant challenges in ensuring robust performance and efficient interference management. Although state-of-the-art multi-user multiple-input multiple-output (MU-MIMO) improves spectral efficiency through spatial multiplexing, its performance is critically dependent on accurate channel state information at the transmitter (CSIT) \cite{Dizdar2021,Xu2022}.

In MU-MIMO systems, rate-splitting multiple access (RSMA) is proposed as an effective strategy to overcome the challenges posed by imperfect CSIT \cite{Clerckx2016, Mishra2021}. In contrast to conventional multiple access techniques like space-division multiple access (SDMA), which treats interference as noise, and non-orthogonal multiple access (NOMA), leaning on interference decoding and cancellation, RSMA splits user messages into common and private parts, providing a more flexible approach \cite{Zhu2024} and a seamless integration with multiantenna operation. While private parts are dealt individually, the common message is decoded by all users. This allows RSMA to achieve a balance between interference suppression and decoding \cite{Ou2023}. Furthermore, depending on the network topology and channel state condition, RSMA can adjust the power and message allocation based on interference levels, naturally transitioning between SDMA and NOMA \cite{Mao2022,Salem2023}. Such flexibility in managing interference makes RSMA more robust than conventional approaches, enabling a better mitigation of multi-user interference \cite{Dai2016,Clerckx2023} even in the presence of imperfect CSIT.

Classical approaches to statistically approximating the signal-to-interference-plus-noise ratio (SINR) in MU-MIMO systems often rely on Gamma distributions \cite{Jaramillo2015}. In particular, Gamma-based approximations under outdated CSIT have been proposed in the context of RSMA-enabled MU-MIMO systems \cite{Dizdar2021,Zhu2024}, providing useful tools for estimating achievable rates and deriving performance bounds. However, \textcolor{black}{these existing frameworks simplify the algebraic tractability by neglecting the cross-correlation terms that emerge between the past channel states and the current channel-aging error vectors. Because the first-order moment of this cross-correlation is identically zero, prior literature implicitly assumed its overall statistical impact to be negligible. As a result}, as will be shown later, these approximations tend to significantly underestimate the SINR variance -- especially in scenarios with a large number of antennas or users, which are precisely the regimes of interest in modern MU-MIMO deployments. 
In this work, we propose an enhanced SINR approximation that addresses these limitations by incorporating previously neglected statistical dependencies, leading to a more accurate characterization of system performance under outdated CSIT without affecting complexity. Unlike competing approaches, the proposed approximation maintains excellent accuracy under a broad spectrum of system settings, encompassing different user densities, antenna array sizes, and quality of CSIT.

\textit{Notation:} $f_{(\cdot)}(\cdot)$ denotes a probability density function (PDF); $a$ is a scalar, $\mathbf{a}$ is a vector, and $\mathbf{A}$ is a matrix. Conjugate transpose of $\mathbf{A}$ is denoted by $\mathbf{A}\hham$. $\overline{a}$ denotes the complex conjugate of $a$.
The expectation operator is denoted as $\mathbb{E}\left\{\cdot\right\}$, and $\mathrm{Var}\{\cdot\}$ denotes variance. $\mathcal{C}\mathcal{N}(\mu ,\sigma^2)$ represents the complex Gaussian distribution with parameters $\mu$ and $\sigma^2$; $\mathcal{G}(D ,\Theta)$ represents the Gamma distribution with scale and shape parameters $D$ and $\Theta$, respectively. The symbol $\sim$ reads as \textit{statistically distributed as}. The symbol $\Re\{\cdot\}$ is the real-part operator.

 \section{System Model}\label{sec-sys}
 We consider a MIMO broadcast channel (BC) configuration, consisting of a transmitter with $N_{\textrm{t}}$ transmit antennas serving $K$ single-antenna users; these are indexed by ${k={1,...,K}}$ and grouped in the set $\mathcal{K}$ with ${N_{\rm{t}} > K}$. Similarly to \cite{Dizdar2021}, and for the sake of generality, we assume that one-layer RSMA is employed for multi-user transmission in the DL.


%

In this scheme, each user’s message is split at the transmitter into a private and a common component, allowing receivers to partially decode interference while treating the remaining interference as noise. Specifically, the message $W_k$ (intended for user $k$) is divided into a private part $W_{\textrm{p}k}$ and a common part $W_{\textrm{c}k}$, for all $k \in \mathcal{K}$. 
The common parts $W_{\textrm{c}k}$ are aggregated into a single common message $W_{\rm c}$, which, along with the $K$ private messages $W_{\textrm{p}k}$, is independently encoded into streams $s_{\rm c}$ and $s_k$, respectively, where $\mathbb{E}\{|s_k|^2\} = 1$. All streams are then linearly precoded prior to transmission. Hence, the transmit signal is written as: 
 \begin{equation} 
 	\mathbf {x}=\sqrt {P \alpha_{\rm c}}\mathbf {p}_{\rm c}s_{\rm c}+ \sum _{k \in \mathcal {K}}\sqrt{P\alpha_{k}}{\mathbf {p}_{k}s_{k}},
 	\label{eq:x}
 \end{equation}
where $P$ denotes the average transmit power, \textcolor{black}{$\mathbf {p}_{\rm c}$ represents the precoding vector for the common message, $\mathbf {p}_k$ the private precoding vector targeted to user $k$,} and $\alpha_{\rm c} \geq 0$, $\alpha_k \geq 0$ are the power allocation coefficients satisfying $\alpha_{\rm c} + \sum_{k \in \mathcal{K}} \alpha_k = 1$. Notably, setting $\sum_{k \in \mathcal{K}} \alpha_k = 1$ effectively disables the common stream, reducing the scheme to conventional MU-MIMO based SDMA. All precoding vectors are assumed to be normalized, i.e., they have unit-norm.

The received signal at user-$k$ is given by:
\begin{equation} y_{k}=\sqrt{\mathcal{L}_k}\mathbf {h}_{k}\hham[m]\mathbf {x}+n_{k}, \quad k\in \mathcal {K},\end{equation}
where $n_k \sim \mathcal{CN}(0,\sigma^2_k)$ denotes the additive white Gaussian noise (AWGN) affecting user $k$, and $\mathcal{L}_k$ represent the propagation losses.
%
We adopt the channel model from \cite{Dizdar2021}, in which the channel vector at time instant $m$ is expressed as
\begin{equation}
    \mathbf{h}_k[m] = \sqrt{\epsilon^2}\, \mathbf{h}_k[m-1] + \sqrt{1 - \epsilon^2}\, \mathbf{e}_k[m],
    \label{eq:h_actual}
\end{equation}
%
%


In this model, $\mathbf{h}_k[m]$ represents the instantaneous channel vector observed at time instant $m$. The channel is assumed to follow a spatially uncorrelated Rayleigh flat-fading model, with i.i.d. entries distributed as $\mathcal{CN}(0,1)$. The temporal correlation coefficient $\epsilon$ follows Jakes' model and is given by $\epsilon = J_0(2\pi f_{\rm D} T)$, where $f_{\rm D} = v f_{\rm c} / c$ is the maximum Doppler frequency, $f_{\rm c}$ is the carrier frequency, $v$ is the user velocity, $c$ is the speed of light, and $T$ denotes the channel instantiation interval. The error vector $\mathbf{e}_k[m]$ also consists of i.i.d. entries distributed as $\mathcal{CN}(0,1)$ and is independent of $\mathbf{h}_k[m-1]$. Due to feedback latency and mobility, the transmitter only has access to the outdated channel vector $\mathbf{h}_k[m-1]$ when computing the precoders at time $m$.

We consider zero forcing (ZF) precoders with some power allocation policy for the private streams, so that the equality $|\mathbf{h}_j\hham[m-1]\mathbf{p}_k| = 0$ holds $\forall j \in \mathcal {K},  j\neq k$. The private precoding matrix is calculated as: 
\begin{equation}
	\mathbf{P}_k = \mathbf{H} (\mathbf{H}\hham \mathbf{H})^{-1},
\end{equation}
where $\mathbf{H} = [\mathbf{h}_1,...,\mathbf{h}_K]$ is a $N_{\rm{t}} \times K$ channel matrix, and $\mathbf{P}_k = [\mathbf{p}_1,...,\mathbf{p}_K]$ is a $N_{\rm{t}} \times K$ precoder matrix. \textcolor{black}{Because the common message must be successfully decoded by all users across isotropic spatial directions, joint multi-user precoding optimization for the common stream is heavily limited, especially when the available CSIT is outdated. Hence, the common precoder is chosen as an isotropic} random beamformer independent of $\mathbf {h}_{k}[m -1] $, $\mathbf {e}_{k}[m]$, and $\mathbf{p}_k$, $\forall k \in \mathcal {K}$.

%


The instantaneous SNRs for the common and private streams at user $k$ are formulated as:
\begin{equation} \gamma _{\textrm{c} k}=\frac {P\mathcal{L}_k\alpha_{\rm c}|\mathbf {h}_{k}\hham[m]\mathbf {p}_{\rm c}|^{2}}{\sigma^2_k +P\mathcal{L}_k\sum _{j \in \mathcal {K}}{\alpha_{j}}|\mathbf {h}_{k}\hham[m]\mathbf {p}_{j}|^{2}},\end{equation}

\begin{equation} \gamma _{\textrm{p} k}=\frac {P\mathcal{L}_k{\alpha_{k}}|\mathbf {h}_{k}\hham[m]\mathbf {p}_{k}|^{2}}{\sigma^2_k + P\mathcal{L}_k\sum _{j \in \mathcal {K}, j \neq k}\alpha_{ j}|\mathbf {h}_{k}\hham[m]\mathbf {p}_{j}|^{2}}.\end{equation}

%


Accordingly, the ergodic achievable rates for the common and private streams are expressed as: 
%
%
%
\begin{align} 
	{R}_{\rm c}=&\mathbb {E}\left \lbrace{ \log _{2} \left ({1 \!+ \! \min _{k\in \mathcal {K}} \left [ \gamma _{\textrm{c} k} \right] }\right) \!\!}\right \rbrace, \\ 
	{R}_{k}=&\mathbb {E}\left \lbrace{ \log _{2} \left ({1 + \gamma _{\textrm{p} k} }\right) }\right \rbrace. \label{eq: R_k}
\end{align}

Finally, the ergodic sum-rate is given by: 
\begin{equation}
\label{eq:esr}
R_{\mathrm {RSMA}}={R}_{\rm c}+\sum _{k=1}^{K}{R}_{k}. \end{equation}
\section{Main results}\label{sec-mainresults}


We rewrite the expression for $	{R}_k$ in \eqref{eq: R_k} as: 

\begin{equation}
	{R}_k = \mathbb{E} \left\{ \log_2 \left( 1 + \mathcal{S}_k  X \right) \right\} -  \mathbb{E} \left\{ \log_2 \left( 1 + \mathcal{S}_k  Z \right) \right\},
\end{equation}
with $\mathcal{S}_k=P\mathcal{L}_k/\sigma_k^2$ having SNR dimensions defined for shorthand notation, and where $X$ and $Z$ are defined as:
\begin{equation}
	X = \sum_{j \in \mathcal{K}}\alpha_{j}|\mathbf {h}_{k}\hham[m]\mathbf{p}_j|^2,
	\label{eq:Xorig}
\end{equation}
\begin{equation}
	\textrm{Z} = \sum_{j \in \mathcal{K}, j \neq k} \alpha_{j}|\mathbf {h}_{k}\hham[m] \mathbf{p}_j|^2.
	\label{eq:Zorig}
\end{equation}

Inserting the definition in \eqref{eq:h_actual} into \eqref{eq:Xorig}, we obtain:
\begin{multline}
\hspace{-4.1mm}X = (1-\epsilon^2) \sum_{j \in \mathcal{K}}\alpha_{j} |\mathbf{e}_k\hham[m]\mathbf{p}_j|^2 + \epsilon^2 \alpha_{k}|\mathbf {h}_{k}\hham[m-1]\mathbf{p}_k|^2 +X_{\rm{CC}},
	\label{eq:Xexpand}
\end{multline}
where 
\begin{equation}
\label{eq:CC}
    X_{\rm{CC}} = 2\sqrt{\epsilon^2(1-\epsilon^2)}\alpha_{k} \Re(\mathbf {h}_{k}\hham[m-1]\mathbf{p}_k\mathbf{e}_k\hham\mathbf{p}_k)
\end{equation}
is the cross-correlation term between the effective channel gain and the projection of the channel error into the direction of the user-$k$. Due to the ZF precoder with uniform power allocation and the nature of the Rayleigh fading channel, the term ${|\mathbf {h}_{k}\hham[m-1]\mathbf{p}_k|^2}\sim$ ${\mathcal{G}(N_{\textrm{t}}-K+1,1)}$.  Following
a similar rationale \textcolor{black}{for the case of $Z$, under ZF precoding the multi-user interference components inside $Z$ reduce to $h_k^H[m]p_j = \sqrt{1-\epsilon^2}e_k^H[m]p_j$ for all $j \neq k$. Since the error vector $e_k[m]$ is statistically independent of the precoding matrix, each individual term $|e_k^H[m]p_j|^2$ follows an i.i.d. exponential distribution, i.e., $\mathcal{G}(1,1)$. Hence, the interference term $Z$ is a weighted sum of independent Gamma random variables, which is approximated by a single Gamma distribution using a second-order moment-matching approximation \cite{Dizdar2021}. Hence, }
we rewrite \eqref{eq:Zorig} as: 
\begin{equation}
    	Z  \sim \mathcal{G} \left(\frac{(1-\alpha_{\textrm{c}}-\alpha_{k})^2}{\sum_{j \in \mathcal{K}, j \neq k} \alpha_{j}^2},\frac{(1-\epsilon^2)\sum_{j \in \mathcal{K}, j \neq k} \alpha_{j}^2}{1-\alpha_{\textrm{c}}-\alpha_{k}}\right).
    \label{eq:Z}
\end{equation}

We note that the cross-correlation term in \eqref{eq:CC} cannot be neglected, as classically done in the related literature \cite{Dizdar2021,Zhu2024}, and must be explicitly considered for a proper approximation. With these previous considerations, the key result in this paper is formalized in the following Lemma, \textcolor{black}{which derives an improved statistical model via a moment-matching approximation.}


\begin{lemma}\label{lemma1}
Let the random variable (RV) $X$ be defined as in \eqref{eq:Xexpand}. Then, $X\approx X_{\rm{G}}$ such that the RV $X_{\rm{G}} \sim \mathcal{G}(D,\Theta)$, with \textcolor{black}{the shape parameter being:}
\begin{equation}
D = \frac{((N_{\rm t}-K+1)\epsilon^2\alpha_{k} + (1-\alpha_{\textrm{c}})(1-\epsilon^2))^2}{(N_{\rm t}-K+1)\epsilon^4\alpha_{k}^2+\sum_{j \in \mathcal{K}} \alpha_{j}^2 (1-\epsilon^2)^2+\mu_k},
	\label{eq:D}
\end{equation}
and \textcolor{black}{the scale parameter of the Gamma distribution:}
\begin{equation}
\Theta = \frac{(N_{\rm t}-K+1)\epsilon^4\alpha_{k}^2+\sum_{j \in \mathcal{K}} \alpha_{j}^2 (1-\epsilon^2)^2+\mu_k}{(N_{\rm t}-K+1)\epsilon^2\alpha_{k} + (1-\alpha_{\textrm{c}})(1-\epsilon^2)}
	\label{eq:Theta}
\end{equation}
where $\mu_k$ is:
\begin{equation}
	\mu_k = \mathbb{E}\left\{X_{\rm{CC}}\right\} =  2\epsilon^2\alpha_{k}^2(1-\epsilon^2)(N_{\rm t}-K+1).
    \label{eq:mu}
\end{equation}

\end{lemma}
\begin{proof}\label{proof-lemma1}
See Appendix \ref{sec:app1}. 
\end{proof}
\textcolor{black}{It should be emphasized that the explicit closed-form expressions for the shape and scale parameters are obtained from the orthogonal properties of ZF precoding, mirroring the baseline environments established in literature \cite{Dizdar2021,Zhu2024}}

\begin{remark}
\label{R1}
    The expressions \textcolor{black}{in} \eqref{eq:D} and \textcolor{black}{in} \eqref{eq:Theta} have been algebraically \textcolor{black}{ arranged to align with those of the moment-matching Gamma distriburion framework used in} \cite{Dizdar2021,Zhu2024}, facilitating \textcolor{black}{a direct visual} comparison \textcolor{black}{of the variance correction term $\mu_k$}. Hence, the effect of neglecting the cross-correlation term in \eqref{eq:CC} is captured by parameter $\mu_k$ in \eqref{eq:mu}, which is the power of $X_{\rm{CC}}$. 
\end{remark}

\begin{remark}
\label{R2}
    Interestingly, the mean of the Gamma distribution $\mathbb{E}\{X_{\rm{G}}\}=D\Theta$ does not depend on $\mu_k$. However, the variance $\mathrm{Var}\{X_{\rm{G}}\}=D\Theta^2$ does. Since $N_{\mathrm{t}}> K$, neglecting $\mu$ implies underestimating the true variance of $X$. This is akin to considering that the fading severity of the equivalent channel $X$ is lower than in practice.
\end{remark}

\begin{remark}
\label{R3}
    The variance of the equivalent channel $X$ is related to the Amount of Fading (AoF) performance metric classically used in communication theory \cite{simon2005digital} to quantify fading severity. Underestimating the variance implies assuming a lower AoF, which leads to exceedingly optimistic estimations of the achievable rates.
\end{remark}


\section{Numerical Results}\label{sec-num}

\begin{figure}
\centering 
\includegraphics[width=.99\columnwidth]{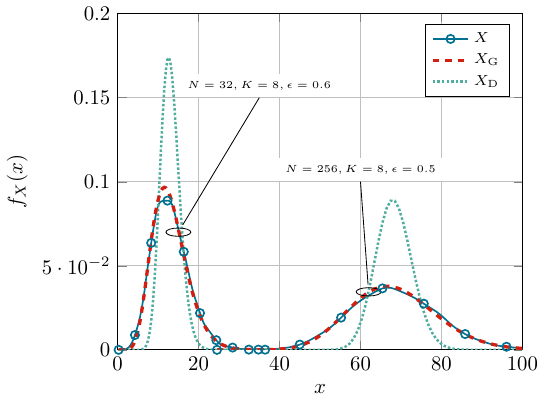}
	\caption{Empirical PDF of $X$ and approximated PDFs of $X_{\rm{G}}$ (Lemma 1) and $X_{\rm{D}}$ \cite{Dizdar2021,Zhu2024}, for different values of $N_{\rm t}$, $K$ and $\epsilon$. 
    }
	\label{fig:XyXest}
\end{figure}

In this section, we validate our analytical results by numerical performance evaluation against Monte Carlo (MC) simulations. First, we focus on providing evidence of the aspects discussed in Remarks \ref{R1} and \ref{R2}. For this purpose, in Fig.~\ref{fig:XyXest} we represent the PDF of $X$ in \eqref{eq:Xexpand}, and provide comparisons with the Gamma approximations in Lemma \ref{lemma1} and \cite{Dizdar2021,Zhu2024}, referred\footnote{Under uniform power allocation, both expressions are formally equivalent.} to in the sequel as $X_{\rm{G}}$ and $X_{\rm{D}}$, respectively. \textcolor{black}{The added correlation parameter $\mu_k$ governs the variance of the PDF. Specifically, neglecting $\mu_k$ (as in $X_{\rm{D}}$) yields an overly narrow and sharp peak in the distribution, which severely overestimates the channel quality.} The cases with $N_{\rm t}=256$, $K=8$ users, CSIT staleness parameter $\epsilon=0.5$, and $N_{\rm t}=32$, $K=8$ and $\epsilon=0.6$ are considered; for simplicity, we use $\alpha_k=1/K$ so that $\mu_k=\mu,\forall k$. As stated in Remark \ref{R2}, the mean value is accurately captured by both approximations, being the first order statistics of the three distributions of interest well-aligned. However, we observe that the Gamma approximations in \cite{Dizdar2021,Zhu2024} largely underestimate the variance; this becomes more noticeable as the excess degrees of freedom captured by $(N_{\rm{t}}-K)$ grows. Conversely, the improved Gamma approximation in Lemma \ref{lemma1} closely matches the true distribution of $X$.

\begin{figure}
\centering 
\includegraphics[width=.99\columnwidth]{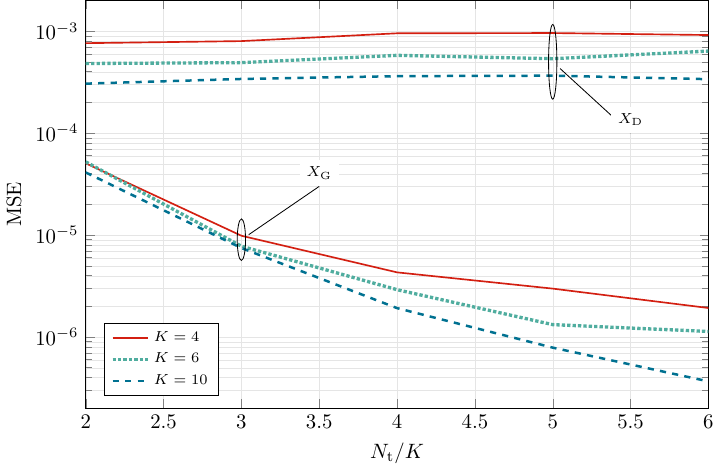}
	\caption{MSE for the Gamma approximations $X_{\rm{G}}$ (Lemma 1) and $X_{\rm{D}}$ \cite{Dizdar2021}, as a function of the ratio $N_{\rm t}/K$ for different values of $K=\{4,6,10\}$. Parameter values are: $\epsilon=0.5$, \textcolor{black}{$M=100$}, and \num{1e5} runs for MC simulations.}
	\label{fig:errorcomb}
\end{figure}

To get further insights into the accuracy of the approximation in Lemma \ref{lemma1}, we evaluate the mean squared error (MSE) between the empirical distribution of $X$ and those corresponding to $X_{\rm{G}}$ and $X_{\rm{D}}$, defined as
\begin{equation}\label{eq1}
    {\mathrm{MSE}}\left(\hat{f}_X,{f}_{X_{\rm{G,D}}};\boldsymbol{\rm \lambda}\right) \triangleq \frac{1}{M}\sum_{k=1}^{M}(\hat{f}_{X}(x_k)-f_{{X_{\rm{G,D}}}}(x_k;\boldsymbol{\rm \lambda}))^2,
\end{equation}
\noindent where $M$ is the number of  \textcolor{black}{sampling} points on which the empirical PDF is estimated, \textcolor{black}{chosen via a uniform grid partioning the continuous support of the empirical PDF, spanning from the minimum to the maximun simulated values of $X$.} ${\boldsymbol{\lambda}}=\left[D,\Theta\right]$ represents the parameters for the corresponding Gamma approximation, and $x_k$ correspond to the abscissa coordinates for such estimations. The MSE is evaluated in Fig. \ref{fig:errorcomb}, as a function of the ratio $N_{\rm t}/K$, considering different numbers of users. We observe that the MSE provided by the approximations in \cite{Dizdar2021,Zhu2024} is consistently larger than that obtained by the approximation in Lemma \ref{lemma1}. While the MSE of the former remains relatively constant as $N_{\rm t}/K$ increases, the MSE of the latter approximation decreases with $N_{\rm t}/K$, with a performance difference over two orders of magnitude. Hence, Lemma \ref{lemma1} provides a useful approximation in the operational regimes of practical MU-MIMO systems, i.e., $N_{\rm t}>>K$.



\begin{figure}
\centering 
\includegraphics[width=.99\columnwidth]{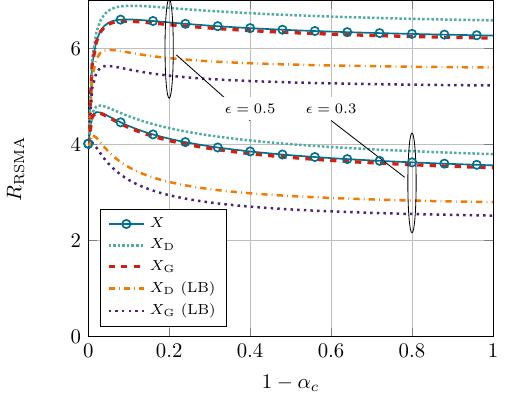}
	\caption{Ergodic sum-rate as a function of the power allocation factor $1-\alpha_{c}$. The exact sum-rate obtained by \eqref{eq:esr}, in solid blue line, is compared to the sum-rates obtained by the proposed Gamma approximation ($X_{\rm G}$) and the existing Gamma approximation ($X_{\rm D}$) in \cite{Dizdar2021,Zhu2024}. The lower bound (LB) proposed in \cite{Dizdar2021} is also evaluated when using ($X_{\rm G}$) and ($X_{\rm D}$) approximations. Parameter values are $N_{\rm t}=16$, $K=4$, with $\epsilon=\{0.3,0.5\}$}
	\label{fig:comparaciones_sumrate}
\end{figure}

Finally, \textcolor{black}{in Fig. \ref{fig:comparaciones_sumrate}} we quantify the impact of using the Gamma approximations in Lemma \ref{lemma1} and \cite{Dizdar2021} in the performance evaluation of a MU-MIMO downlink. 
The ergodic sum-rate in \eqref{eq:esr} is evaluated in different cases: MC simulations implementing $R_{\rm RSMA}$, as well as the approximate rates obtained when using the Gamma approximations for $X_{\rm{G}}$ and $X_{\rm{D}}$ \textcolor{black}{for the private stream. In all instances, we evaluate the common stream rate $R_{\textrm{c}}$ directly from the true sample data.} Parameter values are $N_{\rm t}=16$, $K=4$, $\epsilon=\{0.3,0.5\}$, SNR$=\mathcal{S}_k=20$ dB and $\{\alpha_{k}\}_{k=1}^K = (1-\alpha_{\textrm{c}})/K$. The ergodic sum-rate is evaluated as a function of the RSMA power allocation factor $1-\alpha_{c}$, with the specific case of $1-\alpha_{c}=1$ referring to the reference MU-MIMO based SDMA. As stated in Remarks \ref{R2} and \ref{R3}, the effect of underestimating the true variance in the Gamma approximations in \cite{Dizdar2021,Zhu2024} implies that the corresponding ergodic sum-rate are overestimated. Conversely, we observe that the sum-rate values obtained by the Gamma approximation in Lemma \ref{lemma1} show excellent agreement with the exact values of $R_{\rm RSMA}$ for the entire range of values of $1-\alpha_{c}$. 

There is one side effect of the variance underestimation of the approaches in \cite{Dizdar2021,Zhu2024}: if one wishes to obtain lower bounds for the achievable rates, it is not possible to formally ensure that they always correspond to lower bounds if the starting point is actually an upper bound of the rate. Now, using the lower bound definitions for the achievable rates in \cite[eq. (34)]{Dizdar2021} with the improved values of $D$ and $\Theta$ in our approximation, we see that such bounds are looser than originally envisioned.



\vspace{0cm}\section{Conclusion}\label{sec-con}\vspace{0cm}
We presented a refined Gamma-based approximation for the distribution of the SINR in downlink MU-MIMO systems with outdated CSIT \textcolor{black}{and Zero-Forcing precoders}. The proposed model explicitly accounts for cross-correlation terms typically neglected in prior works, resulting in a more accurate estimation of the SINR variance.

Using an RSMA-based MU-MIMO downlink scenario for benchmarking, we demonstrated that the proposed approximation is particularly effective in massive MIMO regimes (i.e., $N_{\rm t} \gg K$), providing significantly tighter estimates of the ergodic sum-rate and mitigating the optimistic bias introduced by existing models. Future work may extend the proposed analysis to non-Rayleigh fading environments by incorporating different propagation conditions, or to consider alternative precoding schemes such as regularized ZF or MMSE.  


\appendices
\section{Proof of Lemma \ref{lemma1}}\label{sec:app1}

We use a moment-matching approach to approximate $X$ by $X_{\rm{G}} \sim \mathcal{G}(D,\Theta)$, where the shape and scale parameters are computed from the equivalences $D = \mathbb{E}\{X\}^2/\mathrm{Var}\{X\}$, $\Theta = \mathrm{Var}\{X\}/\mathbb{E}\{X\}$, with  $\mathrm{Var}\{X\} = \mathbb{E}\{X^2\} - \mathbb{E}\{X\}^2.$

The first-order moment of $X$ can be calculated as \cite{Zhu2024}:
\begin{equation}
 \mathbb{E}[X] = (N_{\rm t}-K+1)\epsilon^2\alpha_{k} + (1-\alpha_{c})(1-\epsilon^2).  
\end{equation}

Now, the second-order moment requires a lengthy calculation. For convenience, it is expressed as a sum of four components:
\begin{equation}
	\mathbb{E}\{X^2\} = \zeta_1 + \zeta_2 + \zeta_3 + \zeta_4,
\end{equation}
where each term corresponds to:
\begin{align}
	\label{eq:z1}\zeta_1 &=  \mathbb{E}\left\{ \left((1-\epsilon^2) \sum_{j \in \mathcal{K}} \alpha_{j}|\pmb{e}_k\hham[m]\pmb{p}_j|^2\right)^2 \right\}, \\
	\label{eq:z2}\zeta_2 &=  \mathbb{E}\left\{ \left(\epsilon^2 \alpha_{k} |\pmb{h}_k\hham[m-1]\pmb{p}_k|^2\right)^2  \right\}, \\
	\zeta_3 &=  \mathbb{E}\left\{ 2 (1-\epsilon^2)\epsilon^2 \alpha_{k} \sum_{j \in \mathcal{K}} \alpha_{ j}|\pmb{e}_k\hham[m]\pmb{p}_j|^2   |\pmb{h}_k\hham[m-1]\pmb{p}_k|^2 \right\}, \\
	\zeta_4 &=  \mathbb{E}\left\{ \left(2\sqrt{\epsilon^2(1-\epsilon^2)}\alpha_{k} \Re\{\pmb{h}_k\hham[m-1]\pmb{p}_k\pmb{e}_k\hham\pmb{p}_k\}\right)^2 \right\}.
\end{align}

The first two terms can be computed by using the definitions between \eqref{eq:Xorig} and \eqref{eq:CC}, yielding
\begin{align}
	\zeta_1 &= \sum_{j \in \mathcal{K}} \alpha_{j}^2 (1-\epsilon^2)^2 + (1-\epsilon^2)^2(1-\alpha_{c})^2 , \\
	\zeta_2 &= (N_{\rm t} - K + 1)\epsilon^4 \alpha_{k}^2 + (N_{\rm t} - K + 1)^2 \epsilon^4 \alpha_{k}^2.
\end{align}

To derive expressions for $\zeta_3$ and $\zeta_4$, we first perform some algebraic manipulations to obtain (\ref{eq:p3}) and (\ref{eq:p4}), respectively, at the top of next page. In the following derivations, the components of the vectors $\pmb{h}_k$, $\pmb{p}_k$, and $\pmb{e}_k$ are denoted as scalars with an additional subindex. For the sake of compactness, the following ancillary definitions are used in the derivations: $\phi[m] = \mathbb{E}\left\{ \left( \sum _{j=1}^{N_{\rm t}}  h_{k,j}[m-1] \overline{p}_{k,j} \right)^2 \right\}$, and $\xi[m]= \sum _{j=1}^{N_{\rm t}}   \mathbb{E}\left\{ |h_{k,j}[m-1]|^2 |p_{k,j}|^2  \right\}$.

\begin{figure*}[]
\begin{equation}
	\zeta_3 = 2(1-\epsilon^2)\epsilon^2 \alpha_{k}\mathbb{E}\left\{\sum _{j=1}^K \sum _{a=1}^{N_{\rm t}} \sum _{b=1}^{N_{\rm t}} \sum _{c=1}^{N_{\rm t}} \sum _{d=1}^{N_{\rm t}} \alpha_{j} \overline{e}_{k,a}[m] p_{j,a} e_{k,b}[m] \overline{p}_{j,b} \overline{h}_{k,c}[m-1] p_{k,c} h_{k,d}[m-1] \overline{p}_{k,d} \right\},
	\label{eq:p3}
\end{equation}
\begin{equation}
	\zeta_4 = 2(1-\epsilon^2)\epsilon^2 \alpha_{k}^2 \mathbb{E}\left\{\sum _{a=1}^{N_{\rm t}} \sum _{b=1}^{N_{\rm t}} \sum _{c=1}^{N_{\rm t}} \sum _{d=1}^{N_{\rm t}} \overline{h}_{k,a}[m-1] p_{k,a} \overline{e}_{k,b}[m]  p_{k,b} h_{k,c}[m-1] \overline{p}_{k,c} e_{k,d}[m]\overline{p}_{k,d}  \right\},
	\label{eq:p4}
\end{equation}
\hrule
\end{figure*}


\vspace{2mm}
To compute the nested summations in \eqref{eq:p3}, five cases are considered depending on the combinations for the set of summation indices. Specifically, we have $\zeta_3 =  c_1 + c_2 + c_3 + c_4 + c_5$. 

Let us first compute $c_1$, corresponding to the case with $a = b = c = d$. We have:
\begin{multline}
    c_1 = 2(1-\epsilon^2)\epsilon^2 \alpha_{k}\sum _{j=1}^K \sum _{a=1}^{N_{\rm t}} \mathbb{E}\{ \alpha_{j}|e_{k,a}[m]|^2 |p_{j,a}|^2 \\ 
    \times|h_{k,a}[m-1]|^2 |p_{k,a}|^2  \}.
\end{multline}

\textcolor{black}{By the inherent definition of the Gauss-Markov fading process, the channel aging error vector for user $k$ at the time $m$ is statistically independent of the outdated channel vectors of all users $j \in \mathcal{K}$. Because every precoding vector is a deterministic function of these outdated channel vectors, $e_k[m]$ is strictly independent of the precoding components for all users. 
Moreover, the assumption of independence between the user-$k$ components ($h_{k,a}[m-1], p_{k,a}$) and the cross-user precoding entries ($p_{j,a}$ for $j \neq k$) holds asymptotically in the massive MIMO regime ($N_t \gg K$). Due to channel hardening and the isotropic properties of Rayleigh fading, the spatial channels of distinct users are mutually uncorrelated.
With this}, we can reexpress:
\begin{multline}
	c_1 \approx 2(1-\epsilon^2)\epsilon^2 \alpha_{k} \sum _{j=1}^K \sum _{a=1}^{N_{\rm t}} \alpha_{j}\mathbb{E}\left\{ |e_{k,a}[m]|^2 \right\} \mathbb{E}\left\{|p_{j,a}|^2 \right\} \\
	\times\mathbb{E}\left\{ |h_{k,a}[m-1]|^2 |p_{k,a}|^2  \right\}.
\end{multline}

Noting that $\mathbb{E}\left\{ |e_{j,l}[m]|^2 \right\}=1$, $\mathbb{E}\left\{|p_{j,l}|^2 \right\}\approx1/N_{\rm t}$, $\forall j,l$, and $\sum_{j \in \mathcal{K}} \alpha_{j} = (1-\alpha_{c}) $ through mathematical manipulation, we have: 
\begin{equation}
	c_1 \approx 2(1-\epsilon^2)\epsilon^2 \alpha_{k} (1-\alpha_{c}) \phi[m] /N_{\rm t}.
\end{equation}

Following a similar rationale, the cases for $c_2$ to $c_5$ can be computed as summarized in Table \ref{table:cases}, where ${\kappa = 2(1-\epsilon^2)\epsilon^2 \alpha_{k} (1-\alpha_{c})}$.

\begin{table}[]
	\centering
	\caption{Derivation of $\zeta_3$ \eqref{eq:p3}.}
	\begin{tabular}{lll}
		\hline
		Case &             & Expression                                                                                  \\ \hline
		$c_2$ & $a = b = c$       &  $\kappa\left(\phi[m]  - \xi[m]\right)/N_{\rm t}.$  \\
		$c_3$ &$a = b = d$       &  $\kappa \left(\phi[m]  - \xi[m]\right)/N_{\rm t}.$  \\
		$c_4$ &$a = b$ and $c=d$ &  $\kappa (N_{\rm t}-1) \xi[m]/N_{\rm t}.$               \\
		$c_5 $ &$a = b$           & $\kappa (N_{\rm t} - 2) (\phi[m] - \xi[m])/N_{\rm t}.$ \\ \hline
	\end{tabular}
	\label{table:cases}
\end{table}

%
%
%
%
%
%
%

Now, combining all cases together the dependence on the ancillary function $\xi[m]$ vanishes. Hence, we obtain:
\begin{equation}	
	\zeta_3 \approx 2(1-\epsilon^2)\epsilon^2  \alpha_{k} (1-\alpha_{c}) \phi[m]. \label{eq:p3sf}
\end{equation}

Given that the structure of \eqref{eq:p4} is very alike to \eqref{eq:p3}, following a similar procedure of dividing the summation terms in \eqref{eq:p4} into specific cases, we obtain:
\begin{equation}	
	\zeta_4 \approx 2(1-\epsilon^2)\epsilon^2 \alpha_{k}^2 \phi[m].\label{eq:p4sf}
\end{equation}

We see that $\zeta_3$ and $\zeta_4$ are expressed in terms of $\phi[m]$. Because of the structure of the precoding vectors with outdated CSIT, we can safely express: 
\begin{align}
	\phi[m] &= N_{\rm t}-K+1.
    \label{eq:phi}
\end{align}
Finally, combining \eqref{eq:z1}, \eqref{eq:z2}, \eqref{eq:p3}, \eqref{eq:p3sf}, and \eqref{eq:p4sf}, the final expression for the variance is compactly given as:
\begin{equation}
 \mathrm{Var}\left\{X\right\} = (N_{\rm t}-K+1)\epsilon^4\alpha_{k}^2+\sum_{j \in \mathcal{K}} \alpha_{j}^2 (1-\epsilon^2)^2+\mu_k,
\end{equation}
where $\mu_k$ is:
\begin{equation}
	\mu_k = 2\alpha_{k}^2\epsilon^2(1-\epsilon^2)(N_{\rm t}-K+1).
\end{equation}
It is worth noting that the term $\mu_k$ corresponds to the power of the term $X_{\rm{CC}}$. This completes the proof.
\bibliographystyle{IEEEtran}
\bibliography{biblio.bib}

\end{document}